\documentclass[conference]{IEEEtran}
\IEEEoverridecommandlockouts
\usepackage{cite}
\usepackage{amsmath,amssymb,amsfonts} 

\usepackage{algorithmic}
\usepackage{graphicx}
\usepackage{textcomp}
\usepackage{xcolor}
\usepackage{booktabs}
\usepackage{multirow}
\usepackage{multicol}
\usepackage{url}
\usepackage{balance} 
\usepackage{soul}

\usepackage{fancyhdr,lipsum}
\fancypagestyle{firstpage}{
  \fancyhf{}
  \fancyhead[C]{To appear at the 2nd IEEE International Conference on Artificial Intelligence Circuits and Systems (AICAS 2020)}
  \fancyfoot[C]{\thepage}
}

\begin{document}

\title{Using Libraries of Approximate Circuits in Design of Hardware Accelerators of Deep Neural Networks}

\author{\IEEEauthorblockN{Vojtech Mrazek, Lukas Sekanina, Zdenek Vasicek}
\IEEEauthorblockA{Faculty of Information Technology, IT4Innovations Centre of Excellence\\
Brno University of Technology, Brno, Czech Republic \\
\{mrazek,sekanina,vasicek\}@fit.vutbr.cz}
}

\maketitle
\thispagestyle{firstpage}

\begin{abstract}
Approximate circuits have been developed to provide good tradeoffs between power consumption and quality of service in error resilient applications such as hardware accelerators of deep neural networks (DNN). In order to accelerate the approximate circuit design process and to support a fair benchmarking of circuit approximation methods, libraries of approximate circuits have been introduced. For example, EvoApprox8b contains hundreds of 8-bit approximate adders and multipliers. By means of genetic programming we generated an extended version of the library in which thousands of 8- to 128-bit approximate arithmetic circuits are included. These circuits form Pareto fronts with respect to several error metrics, power consumption and other circuit parameters. In our case study we show how a large set of approximate multipliers can be used to perform a resilience analysis of a hardware accelerator of ResNet DNN and to select the most suitable approximate multiplier for a given application. Results are reported for various instances of the ResNet DNN trained on CIFAR-10 benchmark problem.
\end{abstract}


\section{Introduction}


Many computationally intensive applications (such as image recognition, video processing and data mining) feature an intrinsic \emph{error-resilience} property~\cite{chippa}. Since they often process noisy or redundant data and their users are willing to accept certain errors in many cases, the principles of \emph{approximate computing} can be employed in the design of their energy-efficient implementations~\cite{Mittal:2016}. At the circuit level, approximations (i.e. circuit simplifications) are intentionally introduced to find a good trade-off between power consumption, performance and error. A distinguished class of applications among all these error resilient applications are hardware accelerators of deep neural networks (DNNs)~\cite{sze:pieee17}. In the case of DNNs, approximate implementations have been proposed at the level of DNN architecture, data representation, arithmetic operations, memory access and memory cells~\cite{Panda:dnn16,sze:pieee17,Hashemi:Reda:date:2017}.

The approximations can be introduced to the circuit in various steps of the standard circuit design flow. In this work, we primarily focus on the technology independent logic synthesis step. The approximations introduced in this step, the so-called \textit{functional approximations}, modify the Boolean function of the circuit. It has one important advantage —-- the approximate circuit can be implemented in arbitrary ASIC as well as FPGA technology because it is assumed that the technology dependent implementation is performed by means of  common open source or commercial tools after the approximation is conducted.

The methods introduced for the functional approximations can be divided into two categories: (1) manual, and (2) automated. The manual (ad-hoc) methods have been developed for specific circuit components such as adders and multipliers~\cite{Jiang:2017,Mahdiani:TCSI2009}. On the other hand, the automated methods use some general-purpose circuit simplification, resynthesis and approximation techniques and enable us to approximate arbitrary circuits. These methods start with an original (exact) circuit and, typically iteratively, modify its structure.

However, the functional approximation of complex circuits is a time-consuming process. As many of these circuits contain common arithmetic components (circuits) such as adders and multipliers, they can be approximated by replacing selected components by their approximate implementations available in a suitable library.  
A comprehensive library of approximate arithmetic circuits was introduced in 2017, see EvoApprox8b in~\cite{mrazek:date:17}. These circuits were designed by means an automated approximation algorithm based on genetic programming, which will be described in Section~\ref{sec:cgp}. 
EvoApprox8b contains hundreds of 8-bit approximate adders and multipliers. 

The goal of this paper is to extend this library to contain more approximate implementations of arithmetic circuits, with the focus on  hardware accelerators of DNNs. By means of genetic programming we generated an extended version of the library in which thousands of 8- to 128-bit approximate arithmetic circuits are included. These circuits form Pareto fronts with respect to several error metrics, power consumption and other circuit parameters. In our case study we show how a large set of approximate multipliers can be used to perform a resilience analysis of a hardware accelerator of ResNet DNN and to select the most suitable approximate multiplier for a given application. Results are reported for various instances of the ResNet DNN trained on CIFAR-10 benchmark problem~\cite{CIFAR10}.

\section{Automated construction of approximate arithmetic circuits}\label{sec:cgp}

The method used to obtain the library follows the methodology introduced in~\cite{approxBook}. It is a general-purpose approximation method for combinational circuits based on Cartesian Genetic Programming (CGP). CGP represents candidate circuits as directed acyclic graphs and iteratively modifies these circuits to reach the design objectives while ensuring that various constraints (e.g. the error is below a given threshold) are not violated.

\subsection{Errors of approximate circuits}

Selection of the error metrics is the key step of the whole design. The quality of approximate combinational circuits is typically expressed using one or several error metrics, where the most commonly used ones are: the error rate (ER), the mean absolute error (MAE), the mean square error (MSE), the mean relative error (MRE), the worst case error (WCE), the worst case relative error (WCRE), see eq. 1 -- 6 in which the output of the approximate circuit and original (exact) circuit is $O_\mathrm{approx}$ and $O_\mathrm{orig}$, $n_i$ is the number of primary inputs, the operand's width is ${n_i}/2$ bits and $\forall i$ enumerates all possible input vectors.

\begin{align} 
  \mathrm{ER} &= \frac{\sum_{\forall i: O_\mathrm{approx}^{(i)} \neq O_\mathrm{orig}^{(i)}} 1}{2^{n_\mathrm{i}}} \\[0.5em]
  \mathrm{MAE} &= \frac{\sum_{\forall i} \left|O_\mathrm{approx}^{(i)} - O_\mathrm{orig}^{(i)}\right|}{2^{n_\mathrm{i}}} \\[0.5em] 
  \mathrm{MSE} &= \frac{\sum_{\forall i} \left|O_\mathrm{approx}^{(i)} - O_\mathrm{orig}^{(i)}\right|^2}{2^{n_\mathrm{i}}} \\[0.5em]  
  \mathrm{MRE} &= \frac{\sum_{\forall i} \frac{\left|O_\mathrm{approx}^{(i)} - O_\mathrm{orig}^{(i)}\right|}{\max(1, O_\mathrm{orig}^{(i)})}}{2^{n_\mathrm{i}}} \\[0.5em] 
  \mathrm{WCE}  &=\max_{\forall i} \left|O_\mathrm{approx}^{(i)} - O_\mathrm{orig}^{(i)}\right| \\[0.5em] 
  \mathrm{WCRE}  &= \max_{\forall i} \frac{\left|O_\mathrm{approx}^{(i)} - O_\mathrm{orig}^{(i)}\right|}{\max(1, O_\mathrm{orig}^{(i)})} 
\end{align}

\subsection{Cartesian genetic programming}
CGP particularly differs from other genetic programming branches in (1) the solution representation and (2) the search mechanism~\cite{cgp}.

\subsubsection{Representation}
\begin{figure}[ht]
\centering
\includegraphics[width=0.7\linewidth]{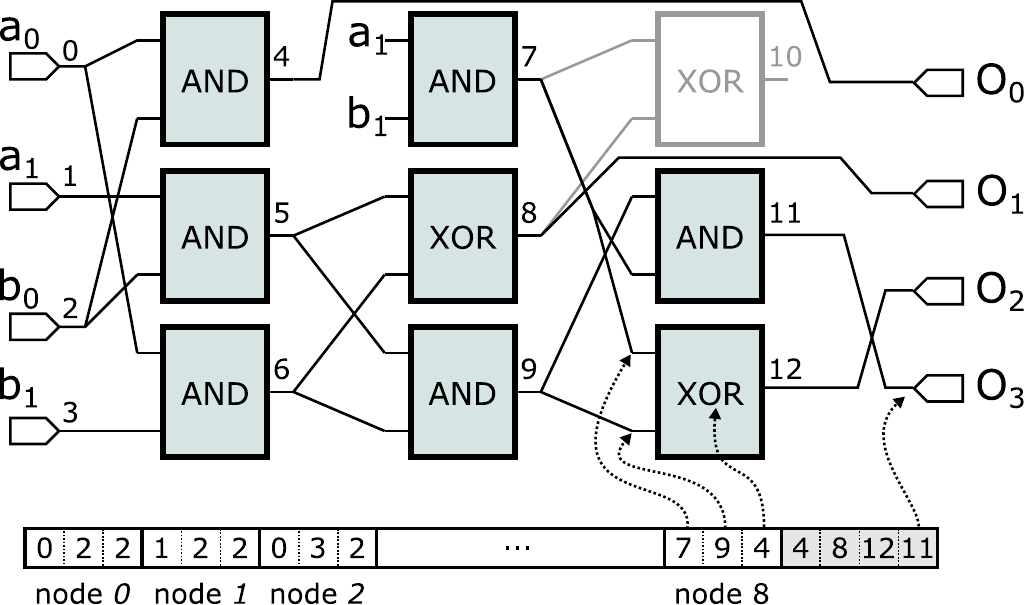}
\caption{A two-bit multiplier represented in CGP with parameters: $n_i = n_o = 4, n_c = n_r = 3, n_a = 2,  \Gamma = \{0^{identity}, 1^{not}, 2^{and}, 3^{or}, 4^{xor}, 5^{nand}, 6^{nor}, 7^{xnor}, 8^{cont0}, 9^{const1} \}$.}\label{fig:cgp}
\end{figure}

A candidate circuit is represented as an integer netlist (the so-called chromosome) describing a constant number of nodes ($N$). These nodes are organized in a two-dimensional grid of $n_c$ columns and $n_r$ rows ($N = n_c \cdot  n_r$). The number of primary inputs and outputs of the circuit is denoted $n_i$ and $n_o$. Each node implements one of the functions specified in the set of functions $\Gamma$ and has up to $n_a$ inputs and a single output. For gate-level circuits, $\Gamma$ usually contains a set of binary logic functions ($n_a = 2$). Fig.~\ref{fig:cgp} gives an example.
 
\subsubsection{Search algorithm}
Every candidate circuit represents one design point in the design space. In CGP, new designs are created by introducing small random modifications to the chromosome. This operation is called the mutation and it typically modifies $h$ integers of the chromosome. Note that all modifications must lead to valid circuits, i.e. only valid function codes and connections can be created.

The search method is based on the $(1+\lambda)$ evolutionary strategy which is usually used for a single-objective circuit approximation~\cite{cgp}. The search algorithm can start with either  a randomly generated initial population or existing designs. The population size is $1+\lambda$. After evaluating the initial population (i.e. measuring the circuit functionality and cost)
the following steps are repeated until the termination condition is not satisfied: (i) the best-scored circuit (called the parent) is selected; (ii) $\lambda$ offspring circuits are created from the parent by means of mutation; (iii) the population is evaluated. 


\subsection{Circuit approximation using CGP}
\label{sec:errororiented}






If a single-objective CGP is applied, the target error range (e.g. the worst-case error), determined by $e_{min}$ and $e_{max}$, is specified by the user. The goal is to minimize the number of gates (or area or power consumption) while the error of the circuits is kept between the target values $e_{min}$ and $e_{max}$. 
If various tradeoffs between the error and the number of gates are requested, CGP is executed several times with $e_{max}$ as the control parameter. 


The multi-objective CGP allows us to optimize the error and other key circuit parameters (area, delay and power consumption) together in one run. We are primarily interested in approximate circuits belonging to the \emph{Pareto front} which contains the so-called \emph{non-dominated solutions}. For example, consider two circuits C1 and C2. Circuit C1 \emph{dominates} circuit C2 if: (1) C1 is no worse than C2 in all objectives, and (2) C1 is strictly better than C2 in at least one objective.


The design process typically starts with an accurate circuit. Candidate approximate circuits are generated from the original circuit using CGP. As millions of candidate circuits are often generated, the evaluation must be fast. For small circuits, an exhaustive circuit simulation utilizing all possible input vectors can be used. However, for large circuits this approach is not scalable. A possible solution is to employ advanced verification methods, based on e.g. Boolean satisfiability solving or binary decision diagram analysis~\cite{ceska:iccad17,vasicek:access2019}. The methodology can easily handle arbitrary constraints.


\section{Library of approximate arithmetic circuits}\label{sec:lib}

Similarly to~\cite{mrazek:date:17} we seeded CGP with conventional implementations of target arithmetic circuits. A typical single-objective CGP run uses the following parameters: $N = k$, where $k$ is the number of gates of the original (exact) circuit with $n_i$ primary inputs and $n_o$ primary outputs, $\lambda = 1$, $h=5$, $\Gamma$ contains all 2-input gates and 1 million generations are produced. One CGP run is typically finished within the order of tens of minutes, depending on the circuit complexity. In the fitness function, the error is obtained by applying one of the error metrics (eq. 1 -- 6) and the cost is estimated as the sum of weighted areas of the gates used in the circuit. At the end of evolution, the best-scored circuit is synthesized and its parameters are determined by a common design tool (we used Synopsys Design Compiler, 45 nm process, $V_{dd}=$1V).  

The new version of the library contains thousands of various arithmetic circuits as shown in Table~\ref{tab:count}. Since the enormous number of circuits makes the selection of the most suitable circuit for a given application difficult, we identified a subset of circuits and used them in our experiments. The selection follows the principles of Pareto optimality with respect to several objectives in which power consumption is compared against EP, MAE, WCE, MSE and MRE metrics. For each of the five subsets of components, ten circuits evenly distributed along the power axis are taken.

\begin{table}[ht]
    \centering
    \caption{The number of approximate implementations of arithmetic circuits in the proposed library}
    \label{tab:count}
    \begin{tabular}{c c r}
    \toprule
         \bf Circuit & \bf Bit-width & \bf \# approx. implementations \\
    \midrule
\multirow{7}{*}{adder} & 8 & 6,979 \\
 & 9 & 332 \\
 & 12 & 4,661 \\
 & 16 & 1,437 \\
 & 32 & 916 \\
 & 64 & 176 \\
 & 128 & 196 \\\midrule
\multirow{4}{*}{multiplier} & 8 & 29,911 \\
& 12 & 3,495 \\
& 16 & 35,406 \\
& 32 & 349 \\\bottomrule
    \end{tabular}
\end{table}

\begin{figure}[ht]
    \centering
    \includegraphics[width=\linewidth]{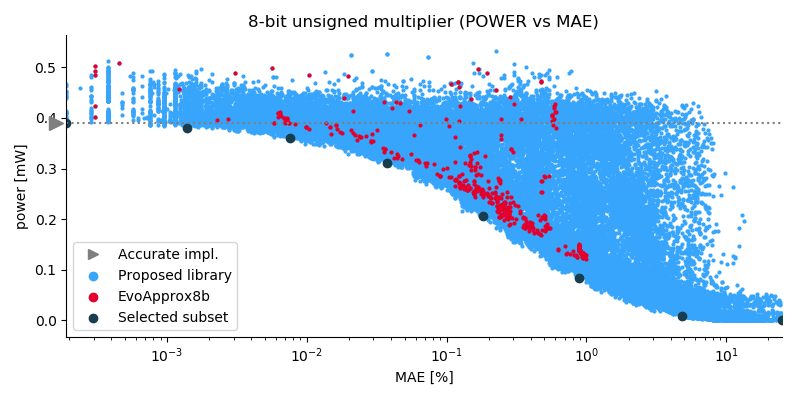}
    \caption{Parameters of 8-bit approximate multipliers (black points) selected from all the approximate multipliers (blue points) and compared to the former version of EvoApprox8b library (red points).}
    \label{fig:cmp}
\end{figure}

Parameters of evolved approximate 8-bit multipliers (power vs. MAE) are shown in Fig.~\ref{fig:cmp}. 
The evolved circuits can be seen as the new state-of-the-art solutions as they provide better tradeoffs than the circuits of the original version of the EvoApprox8b library; the blue points (that occupy the Pareto front ``power vs. MAE'') are clearly better solutions than the red points representing the original circuits of EvoApprox8b. Note that EvoApprox8b was compared with the state of art approximate circuits in a greater detail~\cite{mrazek:date:17,approxBook}. 
Selected approximate circuits and their various parameters can be downloaded from \url{https://ehw.fit.vutbr.cz/evoapproxlib}. 



\begin{table*}[t]
    \centering
    \caption{Parameters of selected approximate multipliers expressed with respect to the exact 8-bit multiplier and the classification accuracy (on CIFAR-10) of various ResNet networks utilizing these circuits. mul8u are evolved multipliers and BAM multipliers ($h$ and $v$ are the horizontal and vertical break levels) are constructed according to~\cite{Mahdiani:TCSI2009}}
    \label{tab:total}
    \resizebox{\textwidth}{!}{
    \begin{tabular}{c|c|ccccc|cccccccc}\toprule
\multirow{2}{*}{\bf Multiplier} & \bf Relative & \multicolumn{5}{c|}{\bf Arithmetic errors}  & \multicolumn{8}{c}{\bf Classification accuracy [\%]} \\
& {\bf Power} [\%] & \it MAE [\%] & \it WCE [\%] & \it MRE [\%] & \it WCRE [\%] & \it ER [\%] & \it ResNet-8 & \it ResNet-14 & \it ResNet-20 & \it ResNet-26 & \it ResNet-32 & \it ResNet-38 & \it ResNet-44 & \it ResNet-50\\\midrule
8 bit (exact) & 100.0  & 0.00 & 0.00 & 0.00 & 0.00 & 0.00 & 82.85 & 85.81 & 88.09 & 89.70 & 88.22 & 89.67 & 88.13 & 89.35\\\midrule
mul8u\_1446 & 99.2  & 0.018 & 0.29 & 0.13 & 28.57 & 9.38 & 82.43 & 85.64 & 88.18 & 89.99 & 87.99 & 89.70 & 88.14 & 89.17\\
mul8u\_2P7 & 98.7  & 0.0015 & 0.0046 & 0.052 & 100.00 & 64.06 & 82.96 & 85.71 & 88.13 & 89.66 & 88.19 & 89.73 & 88.13 & 89.06\\
mul8u\_EXZ & 97.2  & 0.0014 & 0.015 & 0.033 & 28.57 & 19.53 & 82.67 & 85.85 & 88.07 & 89.73 & 88.04 & 89.67 & 88.15 & 89.33\\
mul8u\_KEM & 94.6  & 0.0046 & 0.017 & 0.18 & 100.00 & 75.00 & 82.52 & 85.70 & 88.31 & 89.78 & 88.07 & 89.59 & 88.01 & 89.28\\
mul8u\_GS2 & 91.0  & 0.057 & 1.14 & 0.51 & 64.00 & 29.93 & 82.25 & 85.53 & 88.38 & 89.64 & 88.12 & 89.53 & 88.01 & 88.88\\
m
mul8u\_QJD & 88.0  & 0.017 & 0.082 & 0.51 & 200.00 & 74.80 & 82.61 & 85.99 & 88.17 & 89.96 & 88.32 & 89.39 & 88.19 & 89.12\\
mul8u\_7C1 & 84.1  & 0.13 & 2.38 & 1.04 & 64.00 & 39.93 & 79.95 & 85.00 & 87.67 & 89.87 & 88.18 & 89.28 & 87.89 & 88.82\\
mul8u\_2AC & 79.5  & 0.037 & 0.12 & 1.25 & 3100.00 & 98.12 & 81.37 & 85.59 & 87.70 & 89.81 & 88.23 & 89.36 & 87.55 & 88.29\\
mul8u\_ZFB & 77.7  & 0.059 & 0.45 & 0.80 & 43.56 & 69.26 & 82.03 & 85.76 & 87.96 & 89.63 & 88.28 & 89.37 & 87.49 & 88.01\\
mul8u\_NGR & 70.6  & 0.065 & 0.25 & 1.90 & 150.00 & 96.37 & 81.02 & 85.48 & 88.00 & 89.76 & 88.24 & 89.28 & 87.71 & 88.39\\
mul8u\_PKY & 65.0  & 0.25 & 2.79 & 1.99 & 64.00 & 64.73 & 68.94 & 82.86 & 86.39 & 89.48 & 88.21 & 88.71 & 86.26 & 88.21\\
mul8u\_DM1 & 49.9  & 0.20 & 0.89 & 4.73 & 700.00 & 98.16 & 62.13 & 82.71 & 84.94 & 83.03 & 86.44 & 86.86 & 82.54 & 84.04\\
mul8u\_12N4 & 36.3  & 0.43 & 2.15 & 4.20 & 80.00 & 87.31 & 19.30 & 16.47 & 22.52 & 20.80 & 26.01 & 26.02 & 11.18 & 15.56\\
m
mul8u\_1AGV & 24.3  & 0.67 & 2.94 & 12.14 & 300.00 & 99.05 & 9.12 & 11.82 & 11.53 & 12.96 & 11.58 & 13.14 & 9.74 & 10.39\\
mul8u\_FTA & 21.5  & 0.89 & 4.29 & 13.96 & 125.00 & 98.74 & 8.47 & 13.66 & 11.58 & 9.46 & 10.36 & 12.29 & 12.96 & 11.73\\
mul8u\_YX7 & 15.6  & 4.84 & 49.22 & 15.66 & 66.67 & 88.71 & 12.68 & 10.84 & 10.03 & 9.94 & 10.03 & 10.00 & 10.84 & 11.20\\
mul8u\_JV3 & 8.7  & 2.15 & 8.21 & 39.78 & 7100.00 & 99.16 & 11.10 & 11.05 & 10.99 & 10.89 & 11.24 & 11.48 & 10.31 & 10.42\\
mul8u\_18DU & 7.9  & 2.28 & 9.08 & 28.42 & 100.00 & 99.16 & 9.22 & 10.20 & 9.75 & 9.65 & 10.01 & 10.02 & 11.09 & 10.64\\
\hline
Truncated 7-bit & 75.4  & 0.19 & 0.78 & 2.65 & 100.00 & 74.61 & 48.64 & 77.38 & 72.32 & 60.75 & 78.11 & 79.83 & 62.69 & 64.32\\
Truncated 6-bit & 48.5  & 0.58 & 2.32 & 7.00 & 100.00 & 93.16 & 12.09 & 9.99 & 11.77 & 11.16 & 10.82 & 12.49 & 10.84 & 10.16\\
BAM $h=0, v=2$ & 98.6  & 0.0019 & 0.0076 & 0.077 & 100.00 & 50.00 & 82.75 & 85.66 & 88.23 & 89.77 & 88.19 & 89.66 & 88.18 & 89.18\\
BAM $h=0, v=4$ & 90.6  & 0.019 & 0.075 & 0.56 & 100.00 & 81.25 & 82.86 & 86.07 & 88.19 & 89.90 & 88.34 & 89.74 & 87.97 & 89.20\\
BAM $h=1, v=3$ & 82.2  & 0.10 & 0.40 & 1.47 & 100.00 & 74.80 & 60.61 & 81.02 & 78.41 & 69.86 & 81.83 & 82.93 & 69.60 & 70.72\\
BAM $h=0, v=6$ & 73.9  & 0.12 & 0.49 & 2.64 & 100.00 & 93.75 & 63.51 & 78.60 & 71.27 & 62.92 & 78.57 & 79.25 & 54.21 & 61.39\\
BAM $h=1, v=6$ & 66.2  & 0.20 & 0.78 & 3.43 & 100.00 & 94.34 & 16.34 & 20.80 & 27.26 & 22.81 & 35.20 & 38.79 & 18.05 & 24.78\\
BAM $h=0, v=7$ & 60.5  & 0.29 & 1.17 & 5.21 & 100.00 & 96.48 & 10.39 & 12.16 & 15.01 & 13.67 & 15.74 & 18.17 & 10.62 & 11.81\\
BAM $h=2, v=7$ & 50.2  & 0.49 & 1.95 & 7.00 & 100.00 & 96.97 & 8.90 & 13.63 & 11.05 & 10.62 & 10.04 & 13.15 & 10.23 & 9.39\\
BAM $h=2, v=8$ & 38.7  & 0.78 & 3.13 & 10.56 & 100.00 & 98.14 & 10.13 & 11.29 & 11.53 & 10.06 & 10.09 & 13.35 & 11.30 & 12.43\\

\bottomrule
    \end{tabular}
    }
\end{table*}

\section{Resilience Analysis in DNN Hardware Accelerators}

In order to introduce suitable approximations to DNNs, a resilience analysis is conducted prior to any implementation steps. The resilience analysis of DNNs is usually performed by removing some neurons, weights, memory accesses or inserting some noise to neurons~\cite{Ristretto,Panda:dnn16}. Introducing approximate multipliers to convolutional layers is one of the most preferred approximation techniques~\cite{Sarwar:2018}. The proposed library allows us to perform the resilience analysis (focused on these multipliers) in a more realistic way than previous methods. As many approximate multipliers are available, we can immediately analyze the impact of their utilization not only on the accuracy of classification, but also on the power consumption reduction. 

\begin{figure}[ht]
    \centering
    \includegraphics[width=\columnwidth]{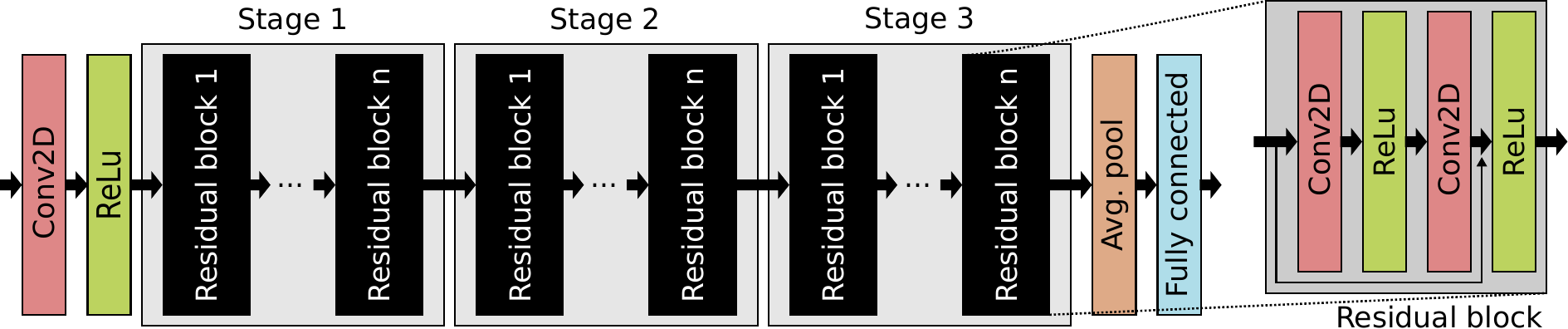}
    \caption{Architecture of ResNet convolutional neural network}
    \label{fig:resnet:arch}
\end{figure}

In our case study, we investigate the impact of approximations introduced to the multipliers used in convolutional layers of ResNet~\cite{ResNet} networks (Fig.~\ref{fig:resnet:arch}) trained to classify CIFAR-10 using TensorFlow~\cite{TensorFlow}. The smallest ResNet-8 network consists of three stages with $n=1$ residual block in each stage. It contains 7 convolutional layers and performs 21 millions multiplications in the inference phase. The classification accuracy drops from 83.42\% to 82.85\% if a common floating point multiplication is replaced with the 8-bit (exact) multiplication. This 8-bit multiplier is considered as a golden solution and all proposed approximations will be compared against it. Note that no retraining is performed after introducing an approximate multiplier.

From the set of 29,911 approximate 8-bit multipliers available in the library, we identified a subset for our experiments in the following way. In order to obtain diverse designs, we selected 10 Pareto optimal multipliers with respect to power and MAE and repeated this selection for other four error metrics. After removing duplicate circuits we ended up with 35 approximate multipliers showing high-quality tradeoffs between power and the five error metrics. 

\begin{figure}[ht]
    \centering
    \includegraphics[width=\columnwidth]{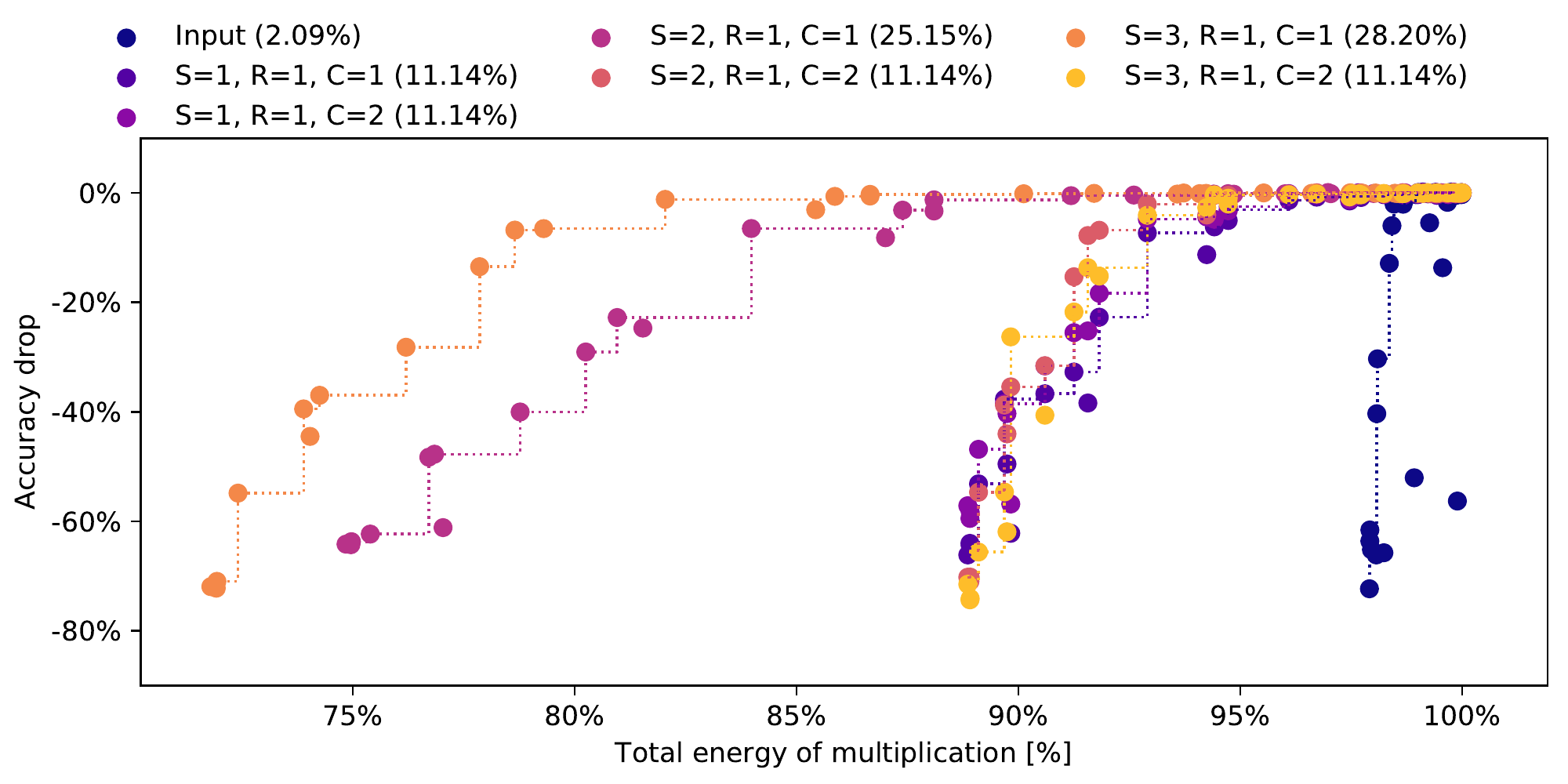}
    \caption{The classification accuracy drop on CIFAR-10 and the power consumption drop measured when approximate multipliers are used in one layer of ResNet-8 (with reference classification accuracy 82.85\%). Different layers are represented using different colors and characterized in terms of the number of stages (S), residual blocks (R), convolutional layers (C) and percentage of multipliers.}
    \label{fig:layerwise}
\end{figure}

All exact multipliers of a given layer of ResNet-8 were then replaced by one of the approximate multipliers. This process has been repeated for all the layers and all the approximate multipliers, but only one layer was modified and one type of approximate multipliers was used in each experiment. Tensor Flow extension TFApprox~\cite{vaverka:date2020} enabled us to accelerate the simulation of ResNet networks containing approximate multipliers. Fig.~\ref{fig:layerwise} shows that the most interesting approximations are obtained if the convolutional layer of the third stage is approximated (see S=3, R=1, C=1 in Fig.~\ref{fig:layerwise}). As this layer contains 28.2\% of all the multipliers, it should undergo the approximation with the highest priority. Introducing the approximate multipliers to the first layer (consisting of 2.09\% multipliers) makes a negligible contribution.

Table~\ref{tab:total} provides a detailed characterization of selected 8-bit approximate multipliers and classification accuracy if these multipliers are employed in all convolutional layers of various instances of ResNet and evaluated on CIFAR-10. Evolved approximate multipliers are compared with common high-quality approximate multipliers created with truncation and BAM algorithm~\cite{Mahdiani:TCSI2009}. One can observe many distinct tradeoffs between classification accuracy and power budget needed for multiplication operations in convolutional layers. This detailed analysis enables the user to identify the best tradeoff for a particular application. For example, if the goal is to reduce power consumption of multipliers to 50\% then it makes no sense to use more complex ResNet than ResNet-32 which shows 86.86\% accuracy.

\section{Conclusions}
In this paper we presented a large library of approximate adders and multipliers which is primarily intended for creating approximate circuits needed in approximate implementations of complex  applications such as energy-efficient hardware accelerators of DNNs. A subset of approximate 8-bit multipliers was then utilized in resilience analysis of ResNet. For various ResNet networks we obtained many tradeoffs between the classification accuracy and power consumption. This knowledge can be exploited during the hardware implementation of DNN accelerator.

Our future work will be devoted to applying the proposed error resilience analysis approach which is based on the existence of a large library of approximate components in other applications.

\vspace{3pt}
\emph{This work was supported by Czech Science Foundation project 19-10137S.} 


\bibliographystyle{./bibliography/IEEEtran}
\bibliography{./bibliography/IEEEabrv,evoapprox}
\end{document}